\documentclass[journal,twocolumn,letterpaper]{IEEEJERM}
\ifCLASSINFOpdf
\else
\fi

\usepackage{times,amsmath,epsfig}
\usepackage{tabularx}
\usepackage{fancyhdr}
\usepackage{amsmath}
\usepackage{amsfonts}
\usepackage{amssymb}
\usepackage[latin1]{inputenc}
\usepackage{float}
\usepackage{array}
\usepackage{graphicx}
\usepackage{url}
\usepackage{subfigure}
\usepackage{bm}
\usepackage{breqn}
\usepackage{xcolor}
\usepackage{soul}
\usepackage{amssymb}
\usepackage{capt-of}
\usepackage{stfloats}
\usepackage{cuted}
\usepackage{multirow}
\begin{document}

\title{U-NET FOR SPECTRAL QUANTITATIVE MICROWAVE BREAST IMAGING}

\author{Ambroise~Di\`es\IEEEauthorrefmark{1},
        H\'el\`ene~Roussel\IEEEauthorrefmark{1}
        and  Nadine~Joachimowicz\IEEEauthorrefmark{1}\IEEEauthorrefmark{2}
        }

\twocolumn[
\begin{@twocolumnfalse}
\maketitle

\begin{abstract}
A spectral approach based on the Fourier diffraction theorem is combined with a pair of U-NETs to perform quantitative microwave imaging of an anthropomorphic breast phantom. The U-NET pair is trained on a spectral database constructed from combinations of different realistic parts of the breast. Some preliminary numerical results are presented to show the major improvement brought by the U-NET.
\end{abstract}

\end{@twocolumnfalse}]

{
  \renewcommand{\thefootnote}{}%
  \footnotetext[1]{\IEEEauthorrefmark{1}Sorbonne Universit\'e, CNRS, G\'enie \'electrique et \'electronique de Paris (GeePs), 75252, Paris, France,\\ ambroise.dies@sorbonne-universite.fr, helene.roussel@sorbonne-universite.fr}
  \footnotetext[2]{\IEEEauthorrefmark{2}Universit\'e Paris Cit\'e, F-75006 Paris France,\\ nadine.joachimowicz@paris7.jussieu.fr}
}
\IEEEpeerreviewmaketitle

\section{Introduction}
\IEEEPARstart{I}{n inverse scattering} for microwave breast imaging, a quantitative description of the breast, corresponding to its dielectric properties, is calculated from the measurement of the scattered field induced by the presence of the breast. From a mathematical point of view, it is an ill-posed problem, due partly to the non-access of evanescent waves, and non-linear because of the multiple scattering effects inside the breast. To solve this, iterative optimization methods incorporating prior information, which minimize the deviation between the calculated and measured scattered field distributions, have been developed. Contrast Source Inversion method (CSI)~\cite{bergContrastSourceInversion1997a}\cite{qinEarlyBreastAnomalies2021}\cite{marianoFieldBasedDiscretization3Contrast2024}, distorted Born Iterative Method (BIM)~\cite{chewReconstructionTwodimensionalPermittivity1990}\cite{neiraHighResolutionMicrowaveBreast2017}, Gauss-Newton~\cite{joachimowiczInverseScatteringIterative1991} or the Broyden-Fletcher-Goldfarb-Shanno method (L-BFGS)~\cite{borzooeiNumericalModelingShoulder2024} are examples of deterministic optimisation algorithm used for various applications. Recent developments in deep learning methods, such as Convolutional Neural Networks (CNNs), are under investigation. These are deterministic methods based on a stochastic optimization process that incorporates an entire database as prior information. Numerous studies have shown that it is possible to improve microwave imaging techniques by combining them with deep learning, as proposed in~\cite{liDeepNISDeepNeural2019}\cite{huangDeepLearningBasedInverse2021a}\cite{zhangUnrolledConvolutionalNeural2023}, and more specifically for the detection of breast tumors~\cite{qinEarlyBreastAnomalies2021}\cite{costanzoFastAccurateCNNBased2023a}\cite{mojabiCNNCompressibilityPermittivity2021}\cite{borghoutsMicrowaveBreastSensing2023}\cite{franceschiniDeepLearningApproach2023}.

This contribution is part of the historical framework of the GeePs-L2S microwave camera~\cite{bolomeyMicrowaveTomographyTheory1990}\cite{henrikssonQuantitativeMicrowaveImaging2010a}\cite{diesAnthropomorphicContourPriori2023}. Previous research on the device is ongoing, focusing on the possibility of training deep learning models to transform the qualitative image produced by the camera, which represents the distribution of induced currents obtained by back-propagating the measured scattered field, into a quantitative image representing the distribution of the object's dielectric properties. The emphasis here is on improving the spectra of induced currents through deep learning.

Contributions that associate deep learning with spectral imaging exist in optics~\cite{chenLearningFullyConnected2022}, but to our best knowledge, no work using a spectral approach has been investigated for quantitative microwave imaging of the breast.

The paper is organized as follows: Section~\ref{sectiondatabase} highlights the database production process, from the anthropomorphic breast model files to the calculation of the truncated spectrum of the  induced currents in the planar configuration of the microwave camera. The optimization process including the architecture of the U-NET and the normalization-denormalization process applied to the data, is detailed in Section~\ref{Section  U-NET}. Finally, selected numerical results, that illustrate both the strengths and limitations of the approach, are shown in Section~\ref{Sectionresultats}.
\section{Database production}
\label{sectiondatabase}
\subsection{Generation of anthropomorphic breast phantoms}

\begin{figure}[!h]
	\centering
	\includegraphics[width=\linewidth*3/4]{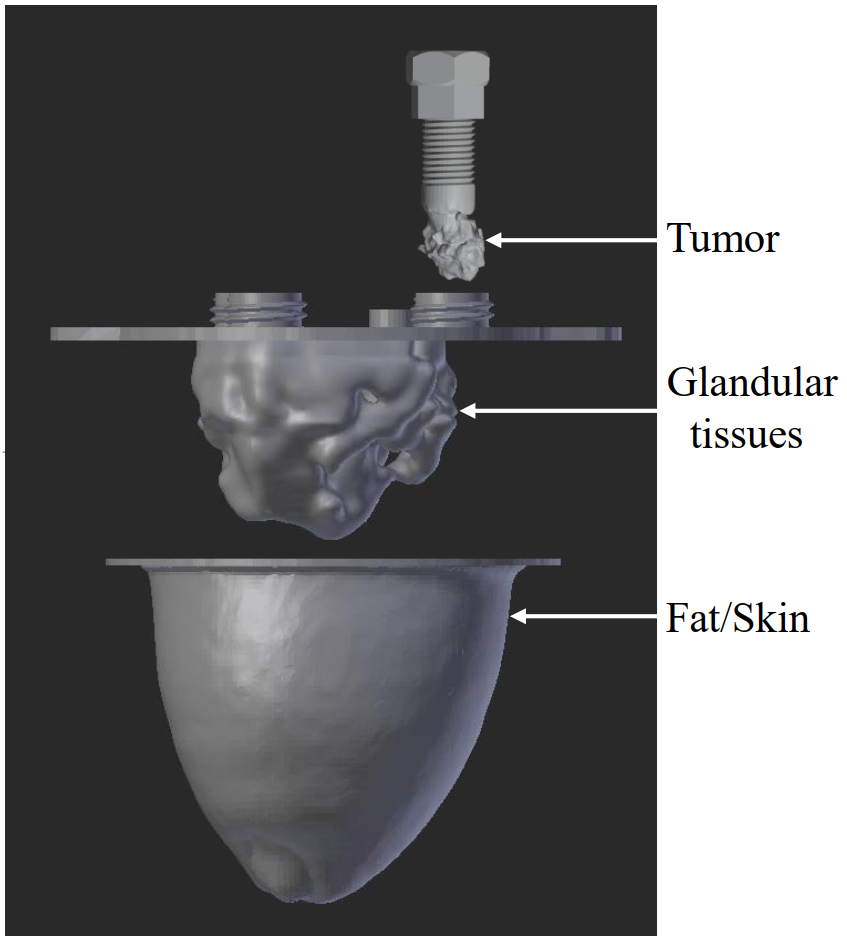}
	\caption{Example of how to combine the different parts of a printable phantom.}
	\label{fig:Combinationparts}
\end{figure}

  To implement deep learning techniques, we created a database based on three-dimensional anthropomorphic models inspired by the GeePs-L2S breast phantom~\cite{joachimowiczAnthropomorphicBreastHead2018}\cite{abediStandardPhantomsEM2022}. The breast is divided into three different parts : skin-associated fat; glandular tissues and tumors as shown in figure~\ref{fig:Combinationparts}. 
We developed, in addition to the GeePs-L2S breast model, three models for fat/skin, five for glandular tissue and five for tumors. This enabled us to produce 56 models without tumors and 168 with tumors, depending on whether the tumor was in the adipose, glandular or both parts. A training set of 1792 images passing through the center of the models was obtained by rotating them, considering 16 different views of each.
For learning purposes, data are randomly separated between a training and a validation dataset in the respected proportion of $80/20\%$ of the database. The original GeePs-L2S breast phantom is reserved for testing purposes. Therefore, the testing set contains 32 spectra.  

\medskip

The dielectric properties of the various parts of the phantom are those provided by standard reference databases of  IFAC \cite{andreuccettiInternetResourceCalculation1997a}, or SUPELEC RECIPES \cite{joachimowiczStandardUwbPhantom}. The values of the relative permittivity, $\epsilon_r$ and conductivity , $\sigma [S/m]$ for each tissue used to perform the simulations, are given at $2.45GHz$ in table~\ref{tab:tabdielectrique}. They include a random variability of $\pm 5\%$.
\begin{table}[H]
\begin{center}
\caption{ $\epsilon_r$ and $\sigma$ for the four tissues of the phantom}
\label{tab:tabdielectrique}
\begin{tabular}{|c|c|c|c|c|}
 \hline
                           & Fat   & Glandular & Skin (Wet) & Tumor\\
 \hline
 $\epsilon_r (\pm 5\%)$    & $5$   & $44$      & $42$       & $53$ \\ 
 \hline
 $\sigma (\pm 5\%)$ [S/m]  & $0.1$ & $1.5$     & $1.6$     & $1.8$\\
 \hline
\end{tabular}
\end{center}
\end{table}
\subsection{Experimental set-up : the GeePs-L2S planar camera}
\begin{figure}[h]
	\centering
	\includegraphics[width=\linewidth*4/5]{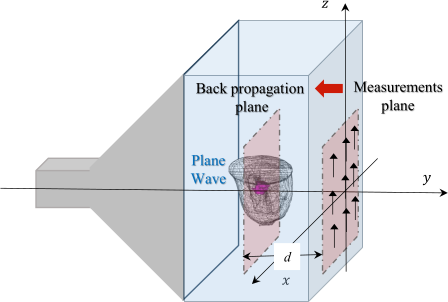}	
	\caption{Configuration of the GeePs-L2S microwave camera.}
	\label{fig:Functionnaldiagram}
\end{figure}

The various breast phantoms in the database are then studied by placing them inside the GeePs-L2S microwave planar camera as illustrated in Figure~\ref{fig:Functionnaldiagram}\cite{henrikssonQuantitativeMicrowaveImaging2010a}. The phantom's model is placed in a water tank whose dielectric properties are  $\epsilon_r = 73$ and $\sigma = 1 [S/m]$. It is illuminated by a monochromatic plane wave polarized along the z axis at the frequency $f=2.45 GHz$. The scattered field is calculated on $64 \times 64$ points of the retina (representing the location of the dipoles) using the commercial electromagnetic software WIPL-D~\cite{WIPLPro2011}, based on a surface integral formulation solved by a method of moments. This constitutes the simulated data for the imaging problem, solved here using a backpropagation algorithm. The observation is placed at $d=10 cm$ from the retina, corresponding to the potential location of a tumor shown in pink in figure~\ref{fig:Functionnaldiagram}.

\noindent We then calculate the filtered spectrum of the induced current at the tumour position using the Fourier diffraction theorem.    
\subsection{Backpropagation projection algorithm}
 The forward electromagnetic problem states that the induced current $\underline{j}$  can be computed by solving the integral equation (\ref{Et_integral_equation}), where $G$ is Green's function;  $\underline{E_t}$ is the total electric field; $k(x,y,z)$ is the wavenumber at the point $(x,y,z)$ and $k_{water}$ is the wavenumber of the coupling medium (water).
\begin{align}
\label{Et_integral_equation}
\underline{E_t}(x,y,z) &= \underline{E_i}(x,y,z) + \left[ \underline{\underline{I}} + \frac{\underline{\nabla} . \underline{\nabla}}{k_{water}^2}\right] .  \nonumber\\
&\iiint_{V} \hspace{1mm} \underline{j}(x',y',z') G(x,y,z;x',y',z')\,dx'\,dy'\,dz'\\
\label{Theoreticaljn}
\underline{j}(x,y,z) &= \left[k^2(x,y,z)-k_{water}^2 \right]\underline{E_t}(x,y,z)
\end{align}
The dipoles placed on the retina of the camera measure the vertical component of the total electric field. We deduce the field scattered by the phantom $\underline{E_s}$ by substracting the incident field along the z axis. Assuming that the depolarization inside the phantom is negligible, it is known that equation (\ref{Et_integral_equation}) can be reduced to a scalar equation. That way, $E_{s,z}$ on the retina, at a distance $d$ of the object, is then written :
\begin{align}
\label{Es_integral_equation}
&E_{s,z}(x,d,z) = \nonumber \\
&\iiint_{V} \hspace{1mm} j_z(x',y',z') G(x,d,z;x',y',z')\,dx'\,dy'\,dz'
\end{align}
Because of Weyl's angular spectrum expansion, Green's function can be expressed as :
\begin{align}
\label{Weylexpansion}
&G(x,d,z;x',y',z') = \nonumber \\
& -\iint_{-\infty}^{+\infty} \hspace{1mm} \frac{j}{8\pi^2 \gamma} e^{-j(k_x(x-x')+\gamma|d-y'|+k_z(z-z'))} dk_x\,dk_z \\
\label{gamma}
&\gamma = \sqrt{k_{water}^2-k_x^2-k_z^2}
\end{align}
If we only consider the propagative waves, equation (\ref{gamma}) enforces that $\gamma \in \mathbb{R}$ and some calculations leads to the Fourier diffraction theorem (\ref{tommodiff})~\cite{devaneyFilteredBackpropagationAlgorithm1982}\cite{kirisitsGeneralizedFourierDiffraction2024a}. It means that the Fourier transform of the induced current $\hat{\j}_z$ is linked to the filtered spectrum, limited to the visible domain, of the scattered field $\hat{E}_{s,z}$. 
\begin{align}
\label{tommodiff}
\hat{E}_{s,z}(k_x,k_z) &= -\frac{j}{2\gamma} \hat{\j}_z(k_x,\gamma -k_{water},k_z)e^{-j \gamma d}
\end{align}

Then, we can easily deduce the backpropagation projection equation (\ref{backpropagationequation})~\cite{bolomeyMicrowaveTomographyTheory1990}, which gives an approximation of the induced current inside the phantom.
\begin{align}
\label{backpropagationequation}
j_z (x,z)= FT^{-1}_{2D} \left[ 2j\gamma  \hat{E}_{s,z}(k_x,k_z)e^{j\gamma d} \right]
\end{align}

\begin{figure}[H]
	\centering
	\includegraphics[width=\linewidth*3/4]{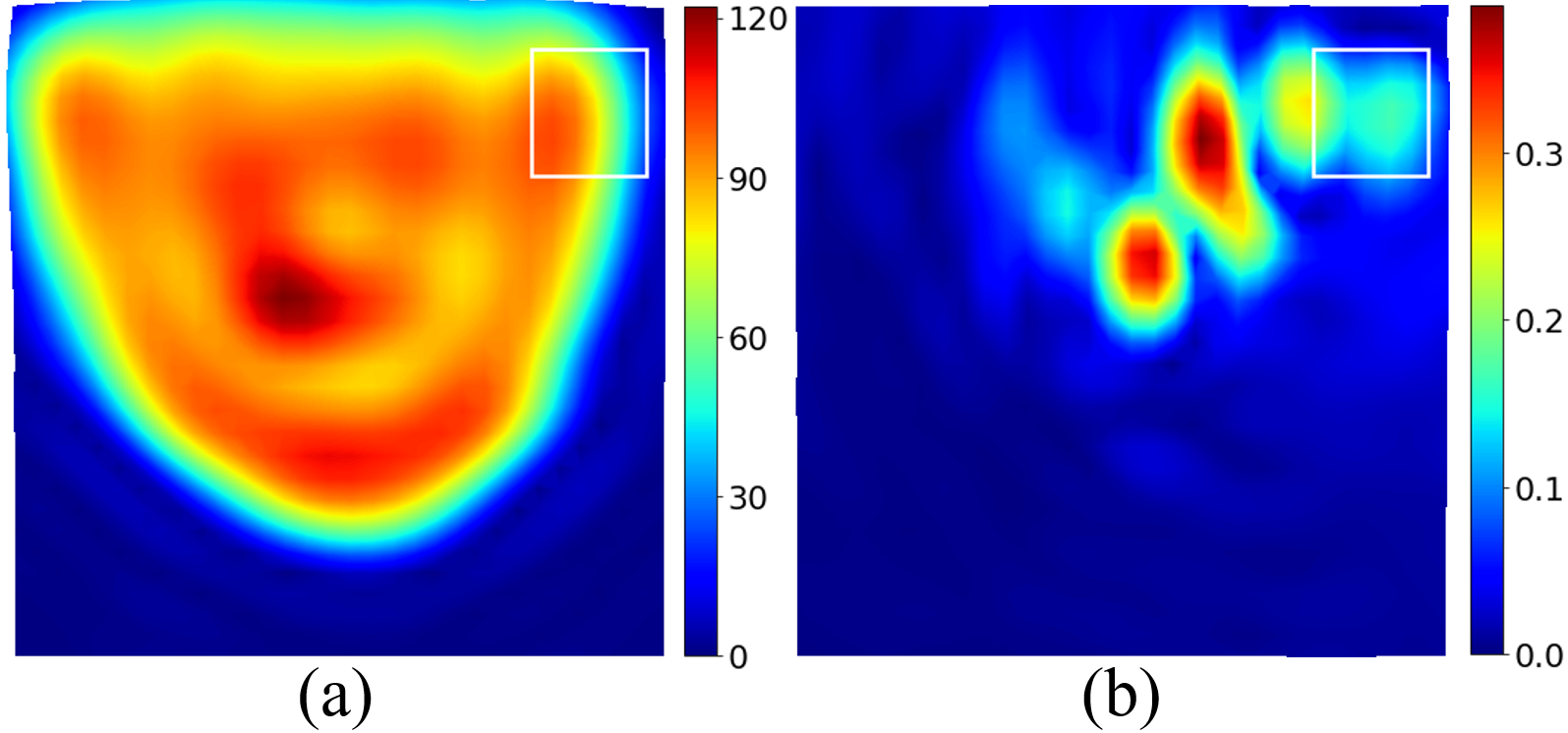}
	\caption{Magnitude of the reconstructed induced current distribution (a); differential image (b). The white rectangle shows the tumor location.}
	\label{fig:Backpropagatedcurrent}
\end{figure}

This formula integrates the induced current of the entire 3D phantom into a single projective plane, making $|j_z|$ given in figure (\ref{fig:Backpropagatedcurrent}.a) difficult to interpret. It also means that the projective plane includes information from the different parts of the phantom, making tumor localization difficult. For assessing the possibility of detecting the tumor, we calculate the difference between the induced current with and without it (\ref{fig:Backpropagatedcurrent}.b). It appears that the tumor information is present in the differential image.

\subsection{Spectral filtering}

\begin{figure}[H]
	\centering
	\includegraphics[width=\linewidth*4/5]{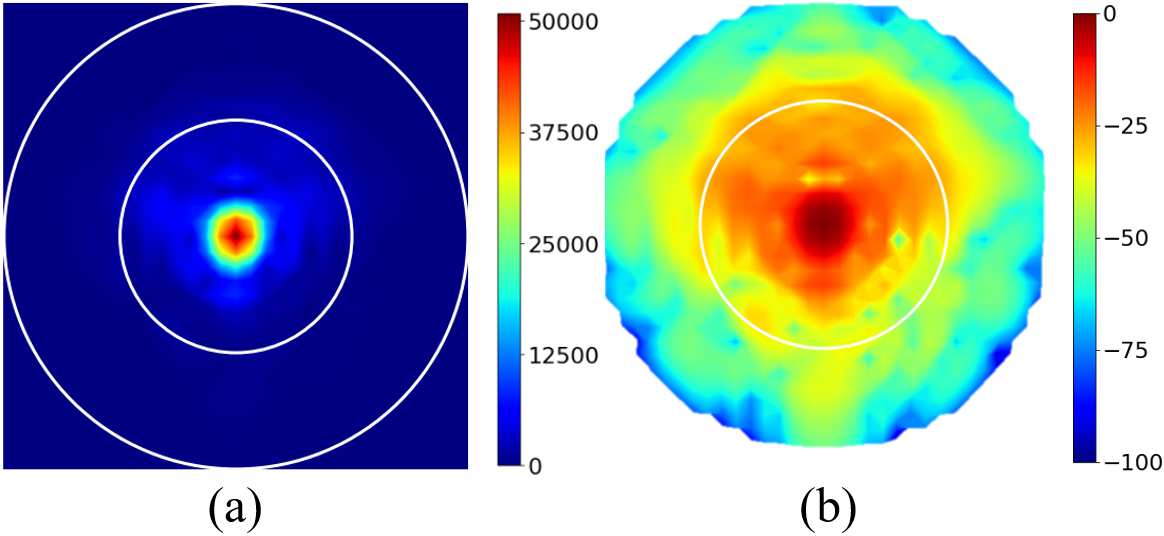}
	\caption{Spectrum (a) and NSD [dB] (b) of the induced current. The white circles represent the visible and half visible domains.}
	\label{fig:Backpropagatedspectrum}
\end{figure}

Once the distribution of scattered field on the retina of the camera  has been obtained, it is possible to deduce its  spectrum using a fast Fourier transform, and then that of these currents using (\ref{tommodiff}). Let's introduce the Normalized Spectral Density of $\hat{\j}_z$, noted NSD, given by :
\begin{align}
	\label{normalized}
	NSD = 10 \ log_{10}\left(\frac{|\hat{\j}_z(k_x,\gamma - k_{water},k_z)|}{max|\hat{\j}_z(k_x,\gamma - k_{water},k_z)|}\right)^2 
\end{align}
In figure \ref{fig:Backpropagatedspectrum} the modulus of $\hat{\j}_z$ distribution of the test phantom and its NSD in decibel [dB] are represented. The two white circles represent the limit of the visible range (outer circle) and this same limit divided by two.

If we examine the modulus of the spectrum, we notice that its level is almost zero outside the inner circle. We can therefore assume that filtering values beyond this circle should not significantly degrade the information.
Spectral filtering offers several advantages. Firstly, spectrum truncation reduces the number of unknowns to be determined, thereby reducing the complexity of the problem posed, from $64 \times 64=4096$ to $197$ pixels.

Secondly, filtering the invisible spectrum and beyond, removes the information contained in evanescent waves, that are not accessible due to measurement noise. This natural truncation in the back-propagation algorithm can be assimilate to a regularization technique, making the spectral technique less sensitive to noise addition. 
\begin{figure}[H]
	\centering
	\includegraphics[width=\linewidth]{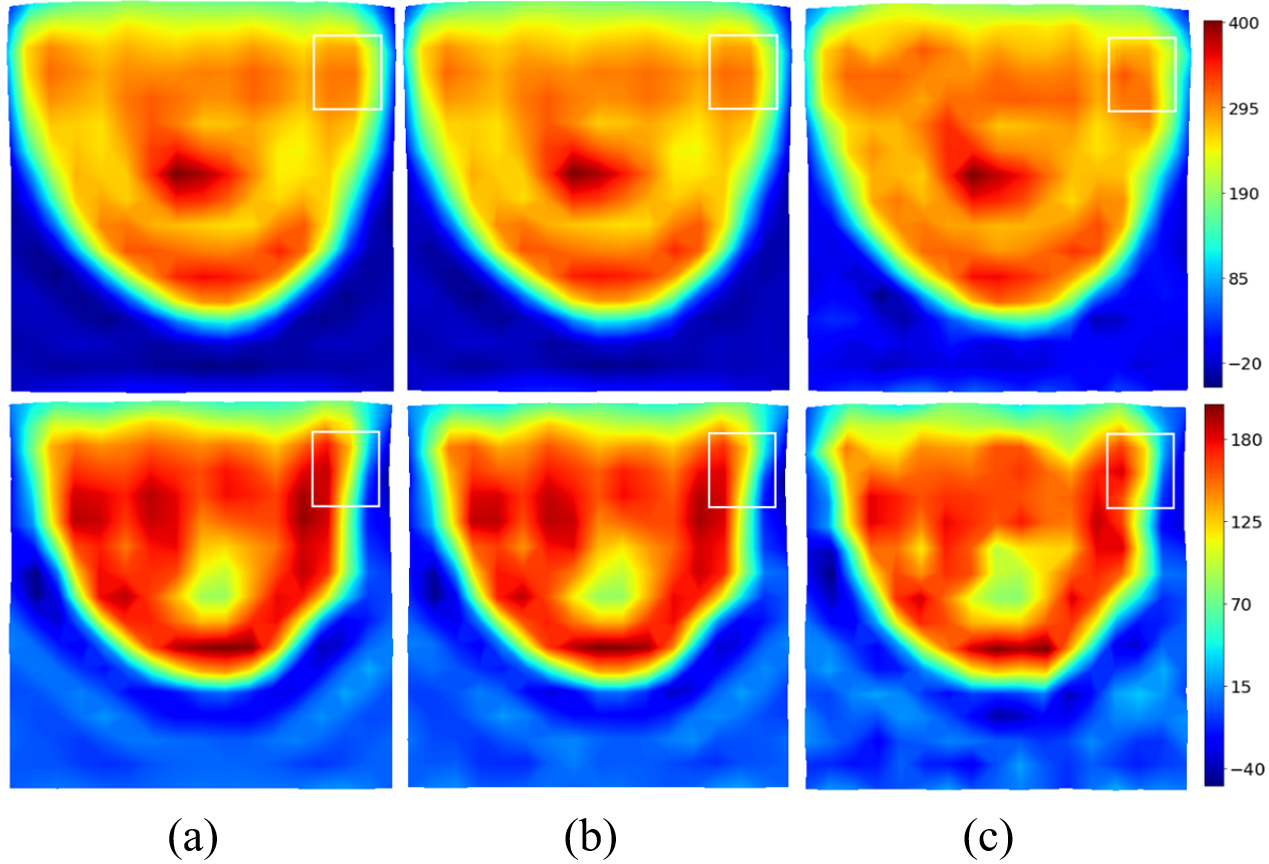}
	\caption{Effect of noise, Top: real part; bottom: imaginary part. Noiseless (a); $SNR_{20dB}$ (b); $SNR_{10dB}$ (c)}
	\label{fig:SNRcomparaison}
\end{figure}
This is illustrated in Figure~\ref{fig:SNRcomparaison}, where the real and imaginary part of the induced currents distributions, have been reconstructed for different signal-to-noise ratios. As expected, the method is very robust, and a signal-to-noise ratio (SNR) of 10 dB has little effect on the reconstruction.

The spectral version of the induced currents, in the visible domain divided by 2, will be taken in the following as the input of the U-NET.
 
\section{U-NET for spectral reconstruction of the dielectric contrast}
\label{Section U-NET}
\subsection{From qualitative to quantitative imaging}
Deep learning techniques are online optimisation processes. To optimize the spectrum of induced currents, a reference must be chosen. The simplest and most logical approach would be to conduct a qualitative-to-qualitative optimization meaning to optimise the approximate induced current with respect to the  theoretical current calculated by the equation (\ref{Theoreticaljn}). However, from a computational viewpoint, this choice is not satisfactory. Considering the variations in the field inside the phantom complicates the process of optimizing the induced current map. The use of the complex contrast $C$, given by equation (\ref{Dielectricontrast}), seems to be a better choice, as it is strongly linked to different parts of the breast and more challenging.
\begin{align}
\label{Dielectricontrast}
C(x,z) &= (\epsilon'(x,z)-\epsilon'_{water}) - j(\epsilon''(x,z) - \epsilon''_{water})\\
\label{epssecsigma}
\epsilon''_{water}&=\frac{\sigma_{water}}{\omega \epsilon_0}
\end{align}

Since the imaginary part of $C$ is small compared its real part, we first rescale both of them into the interval $[0,1]$ by using the standard min-max scaling formula. For the contrast C, this given $C_{min-max}$ :
\begin{align}
C_{min-max} &= \frac{C - min(C)}{max(C) - min(C)}
\label{Minmaxnorm}
\end{align}
As deep learning models are pseudo-probabilistic models, they work best with data close to 0. It is therefore desirable for the real and imaginary parts of the spectrum to lie within the interval $[-1, 1]$. Hence, a normalization is applied to the spectrum $\hat{C}_{minmax}$ according to : 
\begin{equation}
\hat{C}_{norm}(x,z) = \frac{\hat{C}(x,z)_{min-max}}{max(|\hat{C}_{minmax}|)}\\
\end{equation}
The same scaling and normalization process is applied to the spectrum of the filtered induced current $\hat{\j}_{z}$ to $\hat{\j}_{norm}$.

\medskip
The denormalization process that compute $\epsilon_{rnet}$ and $\sigma_{net}$ values from the $C_{net}$ contrast, reconstructed by the U-NET is conducted as follows. 

\noindent Let introduce $M_{\Re}$ and $M_{\Im}$, the mean of the real or imaginary value of all the pixels outside the phantom on the reconstructed image (in water) respectively. As the dielectric properties of water are known, operation (\ref{Calepsr}) and (\ref{Calsigma}) calibrate the reconstructed values, $\epsilon_{cal}$ and $\sigma_{cal}$, with respect to the water, making the process less sensitive to oscillations caused by spectral reconstruction errors. 
\begin{align}
\label{Calepsr}
\epsilon_{cal} &= \left( \frac{\Re(C_{net})}{M_{\Re}} - 1 \right) \epsilon_{r water}\\
\label{Calsigma}
\sigma_{cal}   &= \left( \frac{\Im(C_{net})}{M_{\Im}} - 1 \right) \sigma_{water}
\end{align}
\begin{figure*}[!hb]
	\centering
	\includegraphics[width=\linewidth]{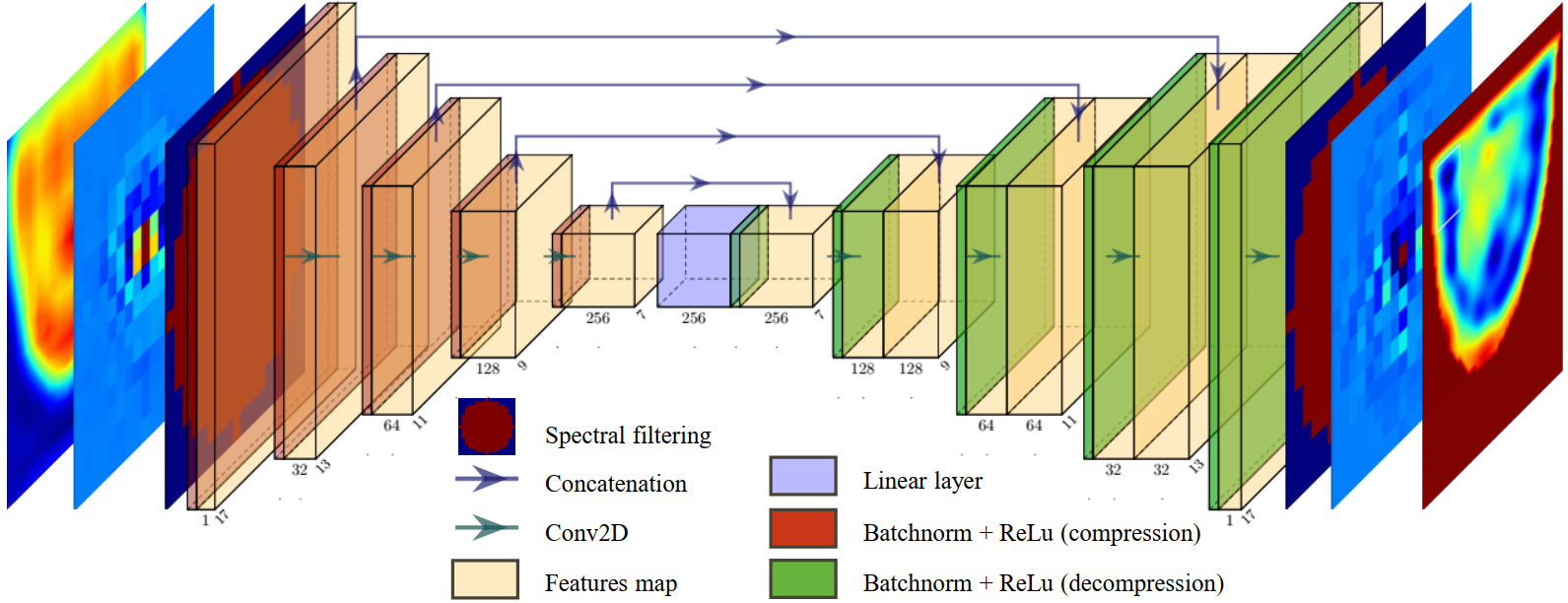}
	\captionof{figure}{U-NET for image enhancement through spectrum transformation}
	\label{fig:U-NET}
\end{figure*}
By using the min-max scaling a second time, we can rescale $\epsilon_{cal}$ and $\sigma_{cal}$ in the interval $\left[min(|\epsilon_{cal}|) ,max(|\epsilon_{cal}|) \right]$ and $\left[min(|\sigma_{cal}|),max(|\sigma_{cal}|) \right]$ to enforce that the maximum and the minimum values reconstructed by the U-NET are positive. That way, equations (\ref{Rescaleeps}) and (\ref{Rescalesigma}) give us the contrast $\epsilon_{rnet}$ and $\sigma_{net}$ reconstructed by the U-NET.
\begin{align}
\label{Rescaleeps}
\epsilon_{rnet} &= min(|\epsilon_{cal}|) + ( max(|\epsilon_{cal}|) - min(|\epsilon_{cal}|)).& \nonumber \\
&\left[ \frac{\epsilon_{cal} - min(\epsilon_{cal})}{max(\epsilon_{cal}) - min(\epsilon_{cal})} \right]\\
\label{Rescalesigma}
\sigma_{net}  &= max(|\sigma_{cal}|) + ( min(|\sigma_{cal}|) - max(|\sigma_{cal}|)). &\nonumber \\ 
&\left[ \frac{\sigma_{cal} - min(\sigma_{cal})}{max(\sigma_{cal}) - min(\sigma_{cal})} \right]
\end{align}

Finally, the application of the contour add prior information by enforcing the constant values of $\epsilon_ {rwater}$ and $\sigma_{water}$ outside the phantom.
\subsection{U-NET scheme}
The model was fully developed hand-made in Python, using functions from the open source PyTorch library~\cite{paszkePyTorchImperativeStyle2019}. Several architectures were tested, including a convolutional neural network, an autoencoder, a modified version of a generative adversarial network and a U-NET. The U-NET proved to be the most effective. For spectrum enhancement, we chose to employ a pair of identical U-NETs~\cite{ronnebergerUNetConvolutionalNetworks2015}, with architecture illustrated in figure~\ref{fig:U-NET}. One U-NET is dedicated to optimizing the real part of the spectrum $\hat{C}_{net}$, while the second U-NET focuses on the imaginary part. The spectral filtering was applied both on the spectral input and output of the U-NET pair to perform a proper pixel-to-pixel transformation of the spectrum.
In the compression phase, we implement a series of filters with decreasing sizes, using zero-padding adjustments to ensure a gradual compression of the spectrum without employing pooling operations. This allows the information to be compressed through convolutions, reducing the dimensions from $17 \times 17$ to $7 \times 7$. The various applied filters gradually generate an increasing number of feature maps, gathering the statistical information needed to modify the spectrum. 
\begin{figure*}[!hb]
	\centering
	\includegraphics[width=\linewidth]{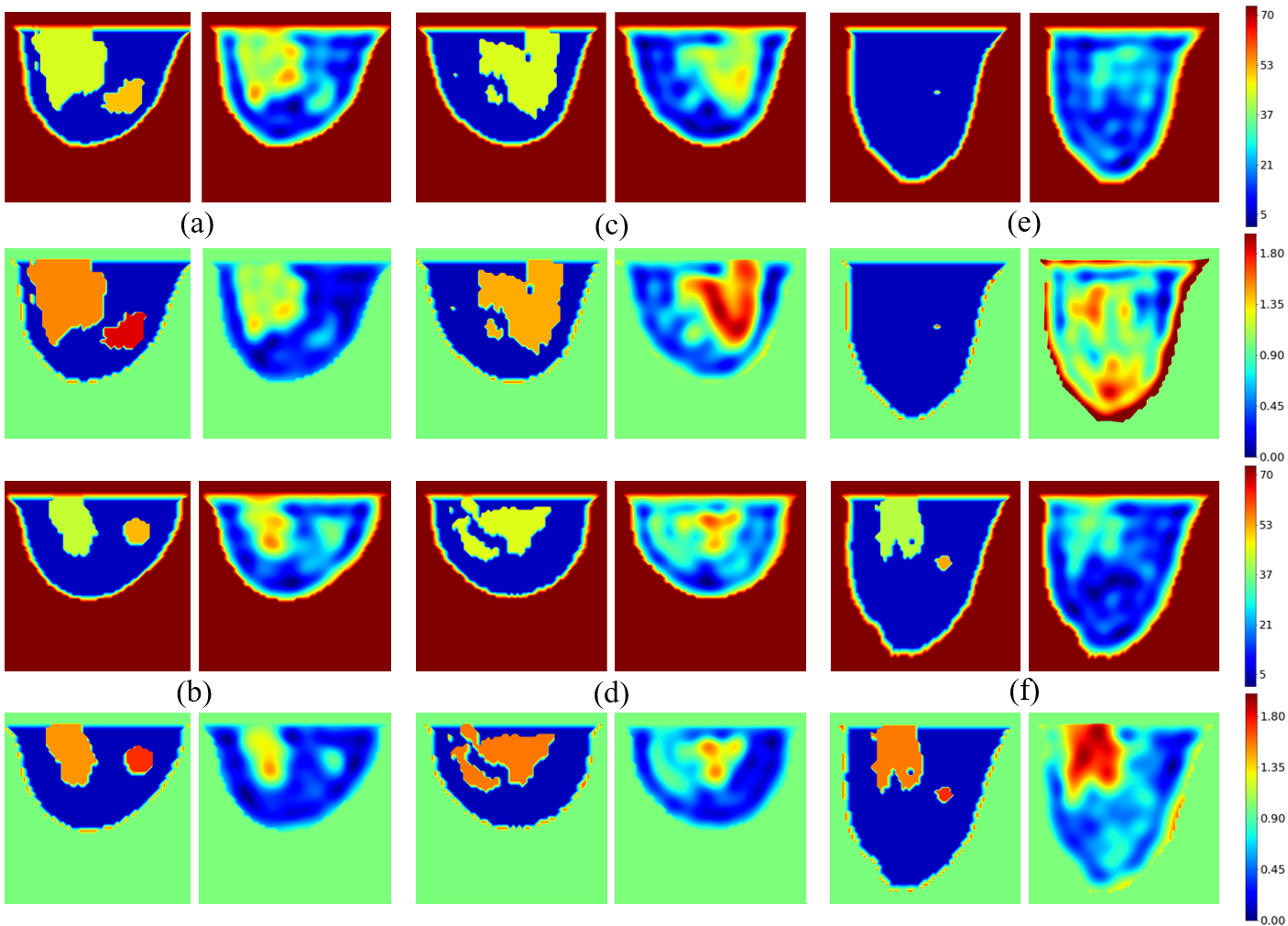}
	\caption{Validation examples : phantoms with tumour (a),(b); phantoms without tumour (c),(d); failure cases (e),(f). The color scale is fixed to $[1,73]$ on $\epsilon_r$ and $[0,2] [S/m]$ on $\sigma$ to fit all the values distant of $\pm 5\%$ from the standard ones.}
	\label{fig:Comparaisonvalidation}
\end{figure*}
During the decompression phase, the filter dimensions are unchanged, and the padding is gradually increased to the initial spectrum size at the end of the network. The sizes of the filters and the corresponding padding for each stage of the U-NET architecture are presented in table~\ref{tab:Filtersizepadding}. 
\begin{table}[H]
\begin{center}
\caption{Filter size and padding}
\label{tab:Filtersizepadding}
\begin{tabular}{|c|c|c|c|c|c|c|c|c|}
\hline
\multirow{1}{*}{ } &
  \multicolumn{4}{c|}{Compression} &
  \multicolumn{2}{c|}{Decompression}\\
\hline
Filter   & 1              & 2             & 3             & 4             & $5 \dots 7 $       & 8 \\
\hline
Size     & $9 \times 9 $  & $7 \times 7 $ & $5 \times 5 $ & $3 \times 3 $ & $3 \times 3 $      & $3 \times 3 $ \\ 
 \hline
Padding  & $ 2 $          & $2$           & $1$           & $0$           & $2$                & $3$\\ 
\hline
\end{tabular}
\end{center}
\end{table}
Before each convolution, batch normalization is applied to speed up the learning process, and the ReLu function is introduced to add non-linearity to the U-NET. The transition from compression to decompression is a linear layer.

Adam optimizer~\cite{kingmaAdamMethodStochastic2017}  minimizes the Weighted Mean Absolute Percentage Error loss (WMAPE) for the real and imaginary components of the spectrum, given by equations (\ref{WMAPEloss1}) and (\ref{WMAPEloss2}), respectively. The optimizer parameters are set to their default values: learning rate $(lr) = 0.001$, and inertial coefficients $(\beta 1, \beta 2) = (0.9, 0.999)$.
\begin{align}
\label{WMAPEloss1}
loss_{\Re } &= \frac{\sum_{i=1}^{n} |\Re(\hat{C}_{net}) - \Re(\hat{C}_{norm})|}{\sum_{i=1}^{n} |\Re(\hat{C}_{norm})|}\\
\label{WMAPEloss2}
loss_{\Im } &= \frac{\sum_{i=1}^{n} |\Im(\hat{C}_{net}) - \Im(\hat{C}_{norm})|}{\sum_{i=1}^{n} |\Im(\hat{C}_{norm})|}
\end{align}
To perform the reconstructions, the spectra were grouped into batches of 3, containing a total of $n = 17 \times 17 \times 3$ parameters. The U-NET were trained with 30 epochs. The computation time for the learning phase is less than an hour and a half, on an NVIDIA RTX A2000 8GB.

\section{Numerical results}
\label{Sectionresultats}
\label{hyperparametres}
 The validation set contains $20\%$ of the data. The U-NET was trained with noise-free data in order to assess its effectiveness on well-controlled data. We will see later that the addition of a 20 dB SNR gives the same results as those obtained without noise.
\begin{figure*}[!hb]
	\centering
	\includegraphics[width=\linewidth*2/3]{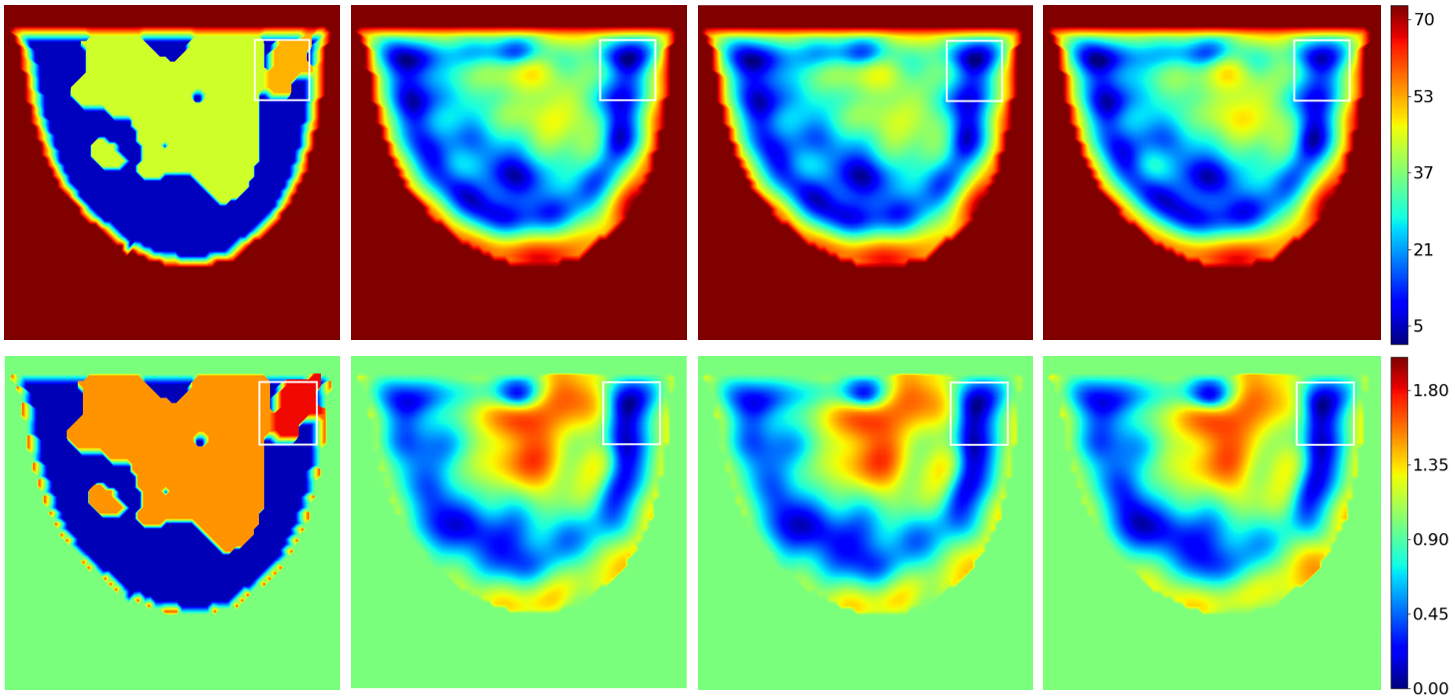}
	\caption{Test example. Top : $\epsilon_r$; bottom :  $\sigma[S/m]$. From the left to the right : Reference; Noiseless reconstruction;  SNR 20dB reconstruction; SNR 10dB reconstruction. The white rectangle shows the tumor location.}
	\label{fig:Comparaisontest}
\end{figure*}
As the models are built from various  combinations of fat/skin, glands and tumors, the validation dataset is not entirely uncorrelated with the training dataset, this generates in our model overfitting issues. It should also be noted that, since the database models are three-dimensional, tumors are not necessarily present in the central plane of observation. The results are shown in figure~\ref{fig:Comparaisonvalidation}. Six models are considered, including three with tumors in the central slice. Four images are presented for each model, the first column corresponds to the exact profile of the dielectric constant (top) and conductivity (bottom), the second column gives the corresponding profiles reconstructed by the U-NET. 

Whatever the model considered, the network generally produces a high-quality image of permittivity and conductivity distributions. The shapes of the various tissues - skin, fat and glands - are well reproduced, with values close to the reference ones. The result is more questionable when the model contains tumors, or more generally for "unusual" situations, leading to failure cases. In the examples of Figure~\ref{fig:Comparaisonvalidation}, the tumor appears in the right place in tumor models~\ref{fig:Comparaisonvalidation}(a),~\ref{fig:Comparaisonvalidation}(b), but not in model~\ref{fig:Comparaisonvalidation}(f) or in the gland-free model ~\ref{fig:Comparaisonvalidation}(e). The reason is probably linked to the training data set, $90\%$ generated from tumor-free breast models in the central breast slice and $5\%$ generated from gland-free models.  To improve these results, the training and validation dataset needs to be expanded. This can be done easily and significantly by adding tumor in the center and diversifying the geometric shapes of the breast models, as this will have a large impact on the total number of possible combinations. Also, with the augmentation of the database, a proper decorrelation between the training and the testing can be performed. These results, obtained with a limited database and a qualitative input image of very low resolution, are very encouraging as preliminary results.

For now, the test set is limited to the GeePs-L2S breast phantom. Figure~\ref{fig:Comparaisontest} shows the dielectric constant and conductivity profiles provided by the network for different signal-to-noise ratios. The effect of noise is almost negligible, confirming the method's low sensitivity to noise for the reasons outlined in section~\ref{Sectionresultats}. If these results are compared with the current distributions (real and imaginary parts) reconstructed by the back-propagation algorithm given in Figure~\ref{fig:SNRcomparaison}, then the improvement is major, both in terms of spatial resolution and of what the image represents. U-NET improves spectrum quality by transforming the low-spatial-resolution qualitative image provided by the backpropagation algorithm into a high spatial resolution quantitative image, giving access to the phantom's dielectric properties. 

\section{Conclusion}
The deep learning approach implemented seems well-suited to the complex problem of retrieving the dielectric properties of an anthropomorphic breast model from the spectrum of induced currents approximated by a sufficiently well-trained neural network.
Preliminary results are impressive, with significant improvements in both image contrast and spatial resolution. However this efficiency will be more effective with a well-dimensioned database, whose development consider the behavior of the neural network in the face of the statistical biases that the database may introduce. Although this database is not currently available, the choice of working in the spectral domain with the backpropagation algorithm seems attractive. The proposed method is robust to noise and easy to implement to evaluate the effectiveness of deep learning. It is potentially a good candidate for a future imaging system.
\section*{Acknowledgment}
The authors would like to thank the Institut Universitaire d'Ing\'enierie en Sant\'e (IUIS) of Sorbonne Universit\'e for financing the doctoral studies. 

\bibliographystyle{IEEEtran}
\bibliography{Article.bib}

\end{document}